\documentclass[aps, prd, 10pt, twocolumn, letterpaper, superscriptaddress, nofootinbib]{revtex4}

\usepackage{amsmath}
\usepackage{amssymb}
\usepackage{mathrsfs}
\usepackage{xspace}
\usepackage[dvipsnames]{xcolor}
\usepackage{braket}
\usepackage{bbold}
\usepackage{graphicx}
\usepackage{nicefrac}
\usepackage{ulem}

\usepackage{ifpdf}
\ifpdf
\fi

\newcommand{\eq}[1]{(\ref{#1})}
\newcommand{\Eq}[1]{Eq.~(\ref{#1})}
\newcommand{\Eqs}[1]{Eqs.~(\ref{#1})}
\newcommand{\Fig}[1]{Fig.~\ref{#1}}
\newcommand{\Sec}[1]{Sec.~\ref{#1}}
\renewcommand{\Ref}[1]{Ref.~\cite{#1}}
\newcommand{\Refs}[1]{Refs.~\cite{#1}}
\newcommand{\App}[1]{Appendix~\ref{#1}}

\newcommand{\ie}{{i.e.,\/}\xspace}
\newcommand{\etal}{{\it et~al.\/}\xspace}

\newcommand{\pd}{\partial}
\newcommand{\del}{\nabla}
\newcommand{\mc}[1]{\mathcal{#1}}
\newcommand{\mcc}[1]{\mathfrak{#1}}
\newcommand{\msf}[1]{\mathsf{#1}}

\newcommand{\oper}[1]{\widehat{#1}}

\renewcommand{\vec}[1]{{\boldsymbol{#1}}}
\newcommand{\favr}[1]{\langle #1 \rangle}
\newcommand{\hm}{a}
\newcommand{\ind}{\diamond}
\newcommand{\m}[1]{\(#1\)}

\newcommand{\ee}{\mathrm{e}}
\newcommand{\ii}{\mathrm{i}}
\newcommand{\dd}{\mathrm{d}}

\renewcommand{\Re}{\text{Re}\,}
\renewcommand{\Im}{\text{Im}\,}

\newcommand{\gc}{G_{\rm N}}

\newcommand{\ampder}{\mcc{R}}

\newcommand{\I}{I}
\newcommand{\total}[1]{\msf{#1}}

\begin{document}

\title{Gravitational wave modes in matter}

\author{Deepen Garg}
\affiliation{Department of Astrophysical Sciences, Princeton University, Princeton, New Jersey 08544, USA}
\author{I. Y. Dodin}
\affiliation{Department of Astrophysical Sciences, Princeton University, Princeton, New Jersey 08544, USA}
\affiliation{Princeton Plasma Physics Laboratory, Princeton, NJ 08543, USA}

\date{\today}

\begin{abstract}
A general linear gauge-invariant equation for dispersive gravitational waves (GWs) propagating in matter is derived. This equation describes, on the same footing, both the usual tensor modes and the gravitational modes strongly coupled with matter. It is shown that the effect of matter on the former is comparable to diffraction and therefore negligible within the geometrical-optics approximation. However, this approximation is applicable to modes strongly coupled with matter due to their large refractive index. GWs in ideal gas are studied using the kinetic average-Lagrangian approach and the gravitational polarizability of matter that we have introduced earlier. In particular, we show that this formulation subsumes the kinetic Jeans instability as a collective GW mode with a peculiar polarization, which is derived from the dispersion matrix rather than assumed a~priori. This forms a foundation for systematically extending GW theory to GW interactions with plasmas, where symmetry considerations alone are insufficient to predict the wave polarization.
\end{abstract}

\maketitle

\section{Introduction}
\label{sec:jeansintro}

The recent observations of gravitational waves (GWs) \cite{ref:abbott16a, ref:abbott16b, ref:abbott17a, ref:abbott17b, ref:abbott17c, ref:abbott17d, ref:abbott19, ref:abbott20a, tex:abbott20b} have boosted interest in basic GW theory. The analytical theory typically focuses, and justifiably so \cite{tex:ligo20, ref:abbott17e, ref:abbott19b, ref:abbott19c}, on GW propagation in vacuum \cite{ref:flanagan05, ref:andersson21}. However, interactions of GWs with gases and plasmas can also be important. For example, in the vicinity of compact GW sources, intense gravitational modes may be able to transfer their energy--momentum to electromagnetic waves or modes strongly coupled with matter, much like how mode conversion works in plasmas \cite{book:tracy, book:stix}. Similarly, the presence of a massive medium in the early Universe could enhance the transfer of the energy--momentum between GWs and the background radiation in ways different from the known photon--graviton conversion \cite{ref:fujita20}. The possible presence of primordial magnetic fields \cite{ref:neronov10, ref:durrer13, ref:subramanian16}, which is now being considered as a significant factor in, for example, recombination \cite{ref:jedamzik20} and Big Bang nucleosynthesis \cite{ref:yamazaki14, ref:luo19}, could also render the collective effects of the plasma important, which are distinct from the fields simply acting as the source term for GWs. Coupling of gravitational and electromagnetic oscillations may also affect how the stochastic GW background influences stimulated emission of electromagnetic radiation in later Universe. Although still hypothetical, these effects are potentially of significant interest, so GW--matter coupling warrants a detailed consideration.

Although GW--matter coupling has been studied in the past, the backreation of matter on metric oscillations is usually ignored (for example, see \Refs{ref:isliker06, ref:brodin00, ref:brodin10b, ref:brodin00b, ref:brodin01, ref:brodin05}) or described in an ad~hoc manner. For example, the polarization of gravitational modes interacting with gases and plasmas is typically assumed either based on symmetry considerations for an isotropic background \cite{ref:asseo76} or simply adopted to be transverse-traceless in line with the vacuum polarizations. But GW--matter coupling within this approximation is weak \cite{ref:flauger18}, so even small additional effects may be important. For example, those could include other polarizations \cite{ref:moretti20}, thermal effects \cite{ref:kumar19}, and fluid viscosity \cite{ref:madore73}. Furthermore, the distortion of the background caused by a local distribution of matter is not necessarily ignorable \cite{foot:sens}. This is a concern particularly because the dispersion and the polarization of waves in matter can evolve when the background parameters evolve \cite{book:tracy, ref:bamba18}. For gravitational modes in particular, this means that the distinction between the tensor modes and other collective oscillations such as Jeans modes becomes blurred \cite{foot:mode}. Hence, a more systematic approach to GWs needs to be developed that would describe \textit{all} linear gravitational perturbations on the same footing without assuming polarization a~priori and include the effects of matter both on the waves and on their background.

Here, we develop such a general formulation using asymptotic methods borrowed from plasma-wave theory \cite{book:tracy, book:stix}. Specifically, we use the standard average-Lagrangian, or Whitham's, approach \cite{book:whitham, ref:dougherty70, ref:dewar77, my:amc, phd:ruiz17}, which allows bypassing the problems associated with covariant self-consistent averaging on curved manifolds \cite{ref:isi18, ref:caprini18, ref:riles13, ref:su12, ref:zalaletdinov96, tex:zalaletdinov97, ref:stein11, ref:green11, tex:green15, ref:buchert15, ref:kaspar12, ref:clarkson11, phd:kaspar14}. (For more details about averaging, see \Ref{tex:mygwquasi}.) In application to GWs, Whitham's approach has been used before \cite{ref:isaacson68a, ref:maccallum73, ref:araujo89, ref:butcher09, my:spinhall, ref:andersson21, tex:mygwquasi}, but here we explicitly employ the average Lagrangian and the gravitational polarizability of matter (specifically, ideal gas, as an example) in oscillating gravitational field, which we introduced earlier in \Ref{my:gwponder}. This approach is convenient in that it shortens the calculations. One can obtain these results using the brute-force approach from \Ref{tex:myql}, but that would require introducing additional machinery that is excessive for the purpose of this work.

We start with the action that describes both the gravitational field and matter. Assuming that the metric perturbations are comprised of small-amplitude quasimonochromatic waves, we simplify this action and then derive a linear equation for these waves in a generic medium. In doing so, we also rederive the second-order component of the Einstein--Hilbert action and compare the expression with the seemingly different ones found in literature \cite{ref:isaacson68a, ref:maccallum73, ref:butcher09, ref:andersson21}. We explore the properties of our GW equation imposed by the requirement of gauge invariance (see also \Refs{tex:mygwquasi, tex:myql, tex:mydecomp}) and study its short-wavelength limit within the geometrical-optics (GO) approximation. 

We show that because matter distorts not only the dispersion of GWs but also the background metric, a consistent GO approximation is possible only when there is an additional (to the inverse wavelength) large parameter, such as the refractive index \m{N}. The usual GWs in dilute plasma have \m{N \sim 1}, so the effect of matter on such waves is comparable to diffraction and thus \textit{must} be neglected within the GO approximation, contrary to what is done usually.  We also consider the ideal-gas model, in which case modes with \m{N \gg 1} become possible. We call these modes gravitostatic by analogy with approximately-electrostatic modes in electromagnetic-dispersion theory \cite{book:stix}. We show that at \m{N \gg 1}, our general relativistic GW equation yields the correct dispersion relation and polarization for the kinetic Jeans instability, which is usually derived separately from the tensor modes \cite{ref:lima02, ref:trigger04, ref:ershkovich08}. These results are intended as a foundation for a future systematic extension of GW theory to GW interactions with plasmas, where coupling with electromagnetic fields must also be accounted for and the wave polarization generally cannot be assumed a~priori or inferred just from symmetry considerations.

This article is organized as follows. In \Sec{sec:jeansprelim}, we introduce the basic concepts and notation. In \Sec{sec:va}, we formulate our variational approach, the GW dispersion operator, and the wave equation for dispersive GWs. In \Sec{sec:go}, we introduce the short-wavelength approximation and discuss gauge invariance of the wave equation. In \Sec{sec:vacuum}, we show how our formulation reproduces the well-known GWs in vacuum. In \Sec{sec:examples}, we introduce the gravitostatic approximation and derive the Jeans-mode dispersion relation and polarization from our GW equation. In \Sec{sec:conc}, we summarize our results. Also, in appendices, we present an explicit derivation of the second-order Einstein--Hilbert Lagrangian density and of the corresponding wave equation.

\section{Preliminaries}
\label{sec:jeansprelim}

\subsection{Einstein equations}
\label{sec:eeq}

Let us consider a metric $\total{g}_{\alpha\beta}$ on a four-dimensional spacetime $x$ with signature $(-+++)$. The dynamics of this metric is governed by the least-action principle \cite{book:landau2}
\begin{gather}\label{eq:S}
\delta S = 0, 
\quad
S = S_{\rm m} + S_{\rm EH}.
\end{gather}
Here, $S_{\rm m}$ is the action of matter (including electromagnetic fields, if any), $S_{\rm EH}$ is the action of the gravitational field called the Einstein--Hilbert action, 
\begin{gather}\label{eq:SEH}
S_{\rm EH} = \frac{1}{2\kappa}\int \dd^4x\,\sqrt{-\total{g}}\,\total{R},
\end{gather}
$\total{R}$ is the Ricci scalar, $\total{g} \doteq \det \total{g}_{\alpha\beta}$, and $\doteq$ denotes definitions. By default, we assume units such that the Einstein constant $\kappa \equiv 8\pi \gc/c^4$ and the speed of light $c$ are equal to unity,
\begin{gather}
c = 8\pi \gc = 1.
\end{gather}

The equations for $\total{g}_{\alpha\beta}$, called the Einstein equations, are obtained from
\begin{gather}\label{eq:Stot}
\frac{\delta S[\total{g}]}{\delta \total{g}^{\alpha\beta}} = 0,
\end{gather}
where $\total{g}^{\alpha\beta}$ is the inverse metric (\(\total{g}^{\alpha\beta}\total{g}_{\beta\gamma} = \delta^\alpha_\gamma\)) and $[\total{g}]$ denotes that the action is evaluated on $\total{g}^{\alpha\beta}$. Using
\begin{subequations}
\begin{gather}
\frac{\delta S_{\rm EH}[\total{g}]}{\delta \total{g}^{\alpha\beta}}
= \frac{\sqrt{-\total{g}}}{2}\,\total{G}_{\alpha \beta},
\label{eq:einsttensor}
\\
\frac{\delta S_{\rm m}[\total{g}]}{\delta \total{g}^{\alpha\beta}} 
= -\frac{\sqrt{-\total{g}}}{2}\,\total{T}_{\alpha \beta},
\end{gather}
\end{subequations}
where $\total{G}_{\alpha \beta}$ is the Einstein tensor and $\total{T}_{\alpha \beta}$ is the local energy-momentum tensor, \Eq{eq:Stot} can be represented~as
\begin{gather}
\label{eq:GT1}
\total{G}_{\alpha\beta} = \total{T}_{\alpha\beta}.
\end{gather}
We will assume, for clarity, that matter is not ultra-relativistic; then $||\total{T}_{\alpha \beta}|| \sim \rho$, where $\rho$ is the mass density.

\subsection{GW and average metric}
\label{sec:jeansmetric}

Let us suppose that $\total{g}_{\alpha\beta}$ can be decomposed as
\begin{gather}
\total{g}_{\alpha\beta} = g_{\alpha\beta} + h_{\alpha\beta},
\label{eq:gtotal}
\end{gather}
where $g_{\alpha\beta}$ has a characteristic magnitude of order one and a characteristic scale $\ell_g$, while $h_{\alpha\beta}$ has a magnitude that does not exceed a small constant $\hm \ll 1$ and a characteristic spacetime scale $\ell_h \ll \ell_g$. More precisely, we assume that there is a scale $\ell_a$ that satisfies
\begin{subequations}\label{eq:golim}
\begin{gather}
\ell_h \ll \ell_a \ll \ell_g,
\label{eq:scalesep}
\\
\epsilon \doteq \ell_h/\ell_g = (\ell_h/\ell_a)^2 = (\ell_a/\ell_g)^2 \ll 1,
\end{gather}
\end{subequations}
and the local average \(\favr{\ldots}\) is introduced over a spacetime volume of size \(\ell_a\). (Various averaging schemes \cite{ref:brill64, ref:zalaletdinov96, tex:zalaletdinov97} can be used to produce equivalent results \cite{ref:isi18, ref:caprini18, ref:riles13, ref:su12,ref:stein11} under the limit of scale separation \eq{eq:scalesep}. For further details about one possible implementation of averaging, see \Ref{tex:mygwquasi}, and a more general approach is presented in \Ref{tex:myql}.) We also assume that any $h_{\alpha\beta}$ of interest is a superposition of quasiperiodic functions, \ie functions of $(\epsilon x, \theta(x))$, where the dependence on $\theta$ is $2\pi$-periodic and has zero average. This entails
\begin{gather}\label{eq:havr}
\favr{h_{\alpha\beta}} = 0.
\end{gather}
We call such a perturbation a GW. Then, \(g_{\alpha\beta}\) can be understood as the background metric for the GW or as the average part of the full metric:
\begin{gather}\label{eq:bgg}
g_{\alpha\beta} = \favr{\total{g}_{\alpha\beta}}.
\end{gather}
This is different from the common approach to linearized perturbative gravity, where an idealized geometry is adopted for the background (usually either the Minkowski metric $\eta_{\alpha\beta}$ or the Friedmann--Lema\^itre--Robertson--Walker metric) and $h_{\alpha\beta}$ absorbs both high-frequency and low-frequency perturbations, namely, $\square \, (h_{\alpha\beta} - \eta_{\alpha\beta}\eta^{\mu\nu}h_{\mu\nu}/2) = -2 \msf{T}_{\alpha\beta}$ in the Lorenz gauge. The problem with this common appraoch is that in the presence of matter, $h_{\alpha\beta}$ can exhibit secular growth, thus invalidating the perturbation approach at large $x$. Our definitions \eq{eq:havr} and \eq{eq:bgg} help avoid this problem and are in line with the standard approach to general-wave problems \cite{book:whitham}.

For any pair of fields $u_1$ and $u_2$ on the background space, we introduce the following inner product:
\begin{gather}\label{eq:inner}
\braket{u_1, u_2} = \int \dd^4 x\,\sqrt{-g}\,u_1^*(x) u_2(x),
\end{gather}
where
\begin{gather}
g \doteq \det g_{\alpha\beta}.
\label{eq:gdet}
\end{gather}
We also introduce the inverse background metric $g^{\alpha\beta}$ via $g^{\alpha\beta}g_{\beta\gamma} = \delta^\alpha_\gamma$, which leads to
\begin{gather}
\total{g}^{\alpha\beta} = g^{\alpha\beta} - h^{\alpha\beta} + {h^\alpha}_\gamma h^{\gamma\beta} + \mc{O}(\hm^3).
\label{eq:hupper}
\end{gather}
Here and further, the indices of the perturbation metric are manipulated using the background metric and its inverse, unless specified otherwise. Also, the sign convention is adopted as in \Refs{book:carroll, book:misner77}, and $\Gamma^\rho_{\mu\nu}$ will denote the Christoffel symbols associated with the background metric:
\begin{gather}
\Gamma^\rho_{\alpha\beta}
= \frac{g^{\rho\sigma}}{2}
\left(
\pd_\alpha g_{\beta\sigma} + \pd_\beta g_{\alpha\sigma} - \pd_\sigma g_{\alpha\beta}
\right).
\label{eq:christoffel}
\end{gather}
The corresponding Riemann tensor is
\begin{gather}
{R^\rho}_{\alpha\sigma\beta}
= \pd_\sigma \Gamma^\rho_{\beta\alpha}
- \pd_\beta \Gamma^\rho_{\sigma\alpha}
+ \Gamma^\rho_{\sigma\lambda} \Gamma^\lambda_{\beta\alpha}
- \Gamma^\rho_{\beta\lambda} \Gamma^\lambda_{\sigma\alpha},
\label{eq:riemann}
\end{gather}
and the Ricci tensor of the background metric is $R_{\alpha\beta} \doteq {R^\rho}_{\alpha\rho\beta}$. We also use the background metric to define the trace-reverse on any given rank-2 tensor \(A_{\alpha\beta}\):
\begin{gather}
\bar{A}_{\alpha\beta} \doteq A_{\alpha\beta} - \frac{1}{2}\, g_{\alpha\beta} A.
\label{eq:trcrvrs}
\end{gather}
Here and further, $A$ denotes the trace of \(A_{\alpha\beta}\) with respect to the background metric, \(A \doteq g^{\alpha\beta}A_{\alpha\beta}\), unless specified otherwise. In particular, $R \doteq g^{\alpha\beta}R_{\alpha\beta}$ and $G \doteq g^{\alpha\beta}G_{\alpha\beta}$, where $G_{\alpha\beta}$ is the background Einstein tensor:
\begin{gather}\label{eq:Gabbg}
G_{\alpha\beta} \doteq \bar{R}_{\alpha\beta} = R_{\alpha\beta} - \frac{1}{2}\, g_{\alpha\beta} R.
\end{gather}
Note that $G_{\alpha\beta}$ can also be expressed as
\begin{gather}
G_{\alpha\beta} =
\frac{2}{\sqrt{-g}}
\frac{\delta S_{\rm EH}[g]}{\delta g^{\alpha\beta}},
\label{eq:backG}
\end{gather}
and the background energy-momentum tensor is
\begin{gather}
T_{\alpha\beta} \doteq
- \frac{2}{\sqrt{-g}}
\frac{\delta S_{\rm m}[g]}{\delta g^{\alpha\beta}}.
\label{eq:backT}
\end{gather}

\section{Variational approach}
\label{sec:va}

\subsection{Basic equations}
\label{sec:beq}

We assume that $g_{\alpha\beta}$ and $h_{\alpha\beta}$ are the only degrees of freedom that describe GWs, meaning that all GW-driven perturbations of matter can be expressed through $h_{\alpha\beta}$. Then, $S[\total{g}]$ can be represented as \cite{tex:mygwquasi}
\begin{gather}\label{eq:redS}
S[g, h]
\approx S[g]
+ S^{(2)}_{\rm m}[g, h]
+ S^{(2)}_{\rm EH}[g, h],
\end{gather}
where the last two terms represent the leading-order GW--matter coupling action and the leading-order GW contribution from the Einstein--Hilbert  action, respectively. Assuming the index notation $\ind \in \{\text{EH}, \text{m}\}$, they can be expressed as
\begin{gather}\label{eq:Sdiam}
S^{(2)}_\ind[g, h] = \frac{1}{2}\braket{h^{\alpha\beta}, \oper{D}^\ind_{\alpha\beta\gamma\delta}h^{\gamma\delta}}.
\end{gather}
If the matter density satisfies \(\rho = \mc{O}(a^2 \ell_h^{-2})\), $S^{(2)}_{\rm m}$ is of the same order of magnitude as the higher-order neglected terms and thus must be neglected as well, relegating the effect of matter only to the background curvature through \(S[g]\). However, under the assumption
\begin{gather}
\rho \gg \hm^2 \ell_h^{-2},
\label{eq:rhoh}
\end{gather}
which is also adopted hereon, corrections to $S[g, h]$ that scale as higher powers of $\hm$ can be neglected without neglecting $S^{(2)}_{\rm m}$. The linear operators $\oper{D}^\ind_{\alpha\beta\gamma\delta}$ that enter \Eq{eq:Sdiam} can be defined~via
\begin{gather}
\oper{D}^\ind_{\alpha\beta\gamma\delta}
h^{\gamma\delta}
\doteq \frac{1}{\sqrt{-g}}\,
\frac{\delta S^{(2)}_\ind[g, h]}{\delta h^{\alpha\beta}},
\label{eq:dm0}
\end{gather}
and they are understood as the gravitational polarizability of vacuum and of matter, respectively. These matrix functions are constrained to satisfy
\begin{subequations}\label{eq:Dsym}
\begin{gather}
(\oper{D}^\ind_{\alpha\beta\gamma\delta})^\dag
=\oper{D}^\ind_{\gamma\delta\alpha\beta},
\label{eq:Dsym1}
\\
\oper{D}^\ind_{\alpha\beta\gamma\delta} 
= 
\oper{D}^\ind_{\beta\alpha\gamma\delta} 
=
\oper{D}^\ind_{\alpha\beta\delta\gamma},
\end{gather}
\end{subequations}
where the dagger denotes Hermitian adjoint with respect to the inner product \eq{eq:inner}. The constraint \eq{eq:Dsym1} reflects the fact that only the adiabatic interactions are captured by the action \eq{eq:redS}. However, it can be waived if an extended variational formulation is used \cite{my:nonloc} or within a more general theory that we do not consider here \cite{tex:myql}. 

The action \eq{eq:redS} leads to the following equation for the perturbation metric \cite{tex:mygwquasi}:
\begin{gather}\label{eq:DX}
(\oper{D}^{\rm m}
+\oper{D}^{\rm EH})_{\alpha\beta\gamma\delta} h^{\gamma\delta} = 0.
\end{gather}
For the background metric, one obtains 
\begin{gather}
G_{\alpha\beta} = T_{\alpha\beta} + \mc{N}_{\alpha\beta},\label{eq:backg}
\end{gather}
where $\mc{N} = \mc{O}(\hm^2)$. In this work, we will assume the linear limit, in which case $\mc{N}$ is negligible. (See \Refs{tex:mygwquasi, tex:myql} for a more general treatment.) Then, with or without coupling to matter, \Eq{eq:DX} can be shown to be invariant with respect to the gauge transformations
\begin{gather}\label{eq:htr}
h^{\alpha\beta} \to h'^{\alpha\beta} = h^{\alpha\beta} -\del^\alpha \xi^\beta - \del^\beta \xi^\alpha,
\end{gather}
where $\xi^\alpha = \mc{O}(\hm)$ is any vector field. This gauge invariance of linearized gravity results from the invariance of the original action \eq{eq:S} with respect to the coordinate transformations $x^\alpha \to x'^\alpha = x^\alpha + \xi^\alpha$. For details and also for an explanation of the gauge invariance beyond the linear approximation, see \Refs{tex:mygwquasi, tex:myql}.

\subsection{Formulas for $\boldsymbol{\oper{D}^\ind_{\alpha\beta\gamma\delta}}$}
\label{sec:S2}

The matter polarizability $\oper{D}^{\rm m}_{\alpha\beta\gamma\delta}$ is generally an integral operator \cite{foot:nonloc}. It can be difficult to calculate without simplifying assumptions, so we postpone discussing it until \Sec{sec:matter}, where an explicit formula for $\oper{D}^{\rm m}_{\alpha\beta\gamma\delta}$ will be presented for a neutral gas within the short-wavelength approximation. In contrast, $\oper{D}^{\rm EH}_{\alpha\beta\gamma\delta}$ can be readily obtained in general, namely, as follows. 

By direct calculation (\App{app:ricciscalar}), we find that the second-order Einstein--Hilbert action can be written as
\begin{gather}\label{eq:2leh}
S^{(2)}_{\rm EH} = \int \dd^4x\,\mc{L}^{(2)}_{\rm EH},
\end{gather}
where \m{\mc{L}^{(2)}_{\rm EH} = \mc{L}^{(2)}_{\rm vac} + \mc{L}^{(2)}_G}, with
\begin{subequations}
\label{eq:Lgrav}
\begin{multline}
\mc{L}^{(2)}_{\rm vac} = \frac{\sqrt{-g}}{4}
\bigg(
- \frac{1}{2}\, \nabla^\rho h^{\alpha\beta} \nabla_\rho h_{\alpha\beta}
+ \frac{1}{2}\, \nabla^\rho h \nabla_\rho h
\\
- \nabla_\alpha h \nabla_\beta h^{\alpha\beta}
+ \nabla^\rho h^{\alpha\beta} \nabla_\alpha h_{\beta\rho}
\bigg),
\label{eq:Lvac}
\end{multline}
\begin{multline}
\mc{L}^{(2)}_G = \frac{\sqrt{-g}}{4}
\bigg(
- \frac{1}{2}\, G h^{\alpha\beta} h_{\alpha\beta}
- G_{\alpha\beta} h^{\alpha\beta} h
\\+ \frac{1}{4}\, G h^2
+ 2 G_{\alpha\beta} h^{\alpha\rho} {h_\rho}^\beta
\bigg).
\label{eq:LG}
\end{multline}
\end{subequations}
(Here \(\nabla\) is the covariant derivative with respect to the background metric and $h \doteq g^{\alpha\beta}h_{\alpha\beta}$ is the trace of the perturbation with respect to the background metric.) This expression is in agreement with those reported in \Refs{ref:isaacson68a, ref:maccallum73, ref:butcher09, ref:andersson21} up to corrections that are important in the context of our article but not in the contexts of the articles mentioned. In particular, \Eq{eq:Lvac} coincides with the expression in Ref.~\cite[Eq.~(5.14)]{ref:isaacson68a}, where $G_{\alpha\beta} = 0$ is assumed and thus $\mc{L}^{(2)}_G = 0$. (The difference in our sign convention and theirs does not affect this result.) Also, the action \eq{eq:2leh} coincides with the one in Ref.~\cite[Eq.~(2.9)]{ref:maccallum73} up to a factor of \(1/2\), which is caused by the difference between our units and theirs leading to their definition of the Einstein--Hilbert action being different from our \Eq{eq:SEH}. Finally, the action assumed in Ref.~\cite[Eq.~(4)]{ref:butcher09} differs from our \Eq{eq:2leh} by factor of \(1/2\) and the last term in the parenthesis in \Eq{eq:LG}. The factor of \(1/2\) comes from a different definition of the action and the energy--momentum tensor. Specifically, the action used in \Ref{ref:butcher09} is twice our $S^{(2)}_{\rm EH}$, but this poses no real problem, because a factor of two is also omitted in the definition of the energy--momentum tensor \eq{eq:backT}. Also note that \Ref{ref:butcher09} assumes a different definition of the background metric [namely, $\total{g}^{\alpha\beta} = \bar{g}^{\alpha\beta} + h^{\alpha\beta}$ instead of our \Eq{eq:gtotal}], which explains why the last term in the parenthesis in \Eq{eq:LG} is missing in Ref.~\cite[Eq.~(6)]{ref:butcher09}. [This also explains the disappearance of the second and the third term from our \Eq{eq:hderivative}.] Finally, \Ref{ref:andersson21} also operates in vacuum without a cosmological constant, leading to $\mc{L}^{(2)}_G = 0$ in their context. Thus the expression in Ref.~\cite[Eq.~(B17)]{ref:andersson21} matches with \Eq{eq:Lvac} upto a constant factor of $1/4$, which is irrelevant in the context of vacuum and is thus omitted there.

By combining \Eqs{eq:dm0} and \eq{eq:2leh}, one obtains
\begin{multline}\label{eq:Dgen}
\oper{D}^{\rm EH}_{\alpha\beta\gamma\delta} h^{\gamma\delta} = \frac{1}{4}\big(
\nabla^\rho \nabla_\rho h_{\alpha\beta}
- g_{\alpha\beta} \nabla^\rho \nabla_\rho h
+ \nabla_\alpha \nabla_\beta h
\\
+  g_{\alpha\beta} \nabla_\rho\nabla_\sigma h^{\rho\sigma}
- \nabla^\rho \nabla_\alpha h_{\beta\rho}
- \nabla_\rho \nabla_\beta {h^\rho}_\alpha
- G h_{\alpha\beta}
\\
- G_{\alpha\beta}h
- g_{\alpha\beta} R_{\rho\sigma} h^{\rho\sigma}
+ 2 G_{\alpha\rho} {h_\beta}^\rho
+ 2 G_{\rho\beta} {h^\rho}_\alpha \big).
\end{multline}
To simplify this expression, we henceforth assume normal coordinates, in which the first-order derivatives of the background metric vanish. The background metric in these coordinates has the form
\begin{gather}
g_{\alpha\beta} = \eta_{\alpha\beta} + \frac{1}{2}\, (\pd_\sigma\pd_\rho g_{\alpha\beta})\, x^\rho x^\sigma + \mc{O}(\ell_g^{-3}),
\label{eq:normal}
\end{gather}
with its double derivatives given by \cite{ref:brewin98}
\begin{gather}
\pd_\sigma \pd_\rho g_{\alpha\beta}
= - \frac{1}{3} \left(
R_{\alpha\rho\beta\sigma} + R_{\alpha\sigma\beta\rho}
\right).
\label{eq:gder}
\end{gather}
Then, $\oper{D}^{\rm EH}_{\alpha\beta\gamma\delta} = (\oper{D}^{\rm vac} + \oper{\mc{G}})_{\alpha\beta\gamma\delta}$, where the operator $\oper{D}^{\rm vac}_{\alpha\beta\gamma\delta} = \mc{O}(1)$ is given by
\begin{multline}
\oper{D}^{\rm vac}_{\alpha\beta\gamma\delta} h^{\gamma\delta} =
\frac{1}{4}\big(
\pd^\rho \pd_\rho h_{\alpha\beta}
- g_{\alpha\beta} g^{\rho\sigma}\pd^\lambda \pd_\lambda h_{\rho\sigma}
+ g^{\rho\sigma} \pd_\alpha \pd_\beta h_{\rho\sigma}
\\
+ g_{\alpha\beta} \pd^\rho \pd^\sigma h_{\rho\sigma}
- \pd^\rho \pd_\alpha h_{\beta\rho}
- \pd^\rho \pd_\beta h_{\alpha\rho} \big)
\label{eq:dvac}
\end{multline}
and $\oper{\mc{G}}_{\alpha\beta\gamma\delta}$ is given by
\begin{multline}
\oper{\mc{G}}_{\alpha\beta\gamma\delta}h^{\gamma\delta} =
\frac{1}{4}\big(
- G h_{\alpha\beta}
- G_{\alpha\beta} h
+ 2 G_{\alpha\rho} {h_\beta}^\rho
\\+ 2 G_{\rho\beta} {h^\rho}_\alpha
- 2 g_{\alpha\beta} R_{\rho\sigma} h^{\rho\sigma}
+ 2 R_{\rho\alpha\sigma\beta} h^{\rho\sigma}
\big).
\label{eq:mcG}
\end{multline}
(An alternative derivation of the above two equations is presented in \App{app:riccitensor}.) The background Riemann tensor $R_{\rho\alpha\sigma\beta}$ can be further expressed through $R_{\alpha\beta}$ and the background Weyl tensor \(C_{\gamma\mu\delta\nu}\) \cite[Sec.~6.7]{book:weinberg}:
\begin{multline}
R_{\rho\alpha\sigma\beta}
=\frac{1}{2}
\left(
g_{\rho\sigma} R_{\alpha\beta}
- g_{\rho\beta} R_{\alpha\sigma}
- g_{\alpha\sigma} R_{\rho\beta}
+ g_{\alpha\beta} R_{\rho\sigma}
\right)
\\
- \frac{R}{6}
\left(
g_{\rho\sigma} g_{\alpha\beta}
- g_{\rho\beta} g_{\alpha\sigma}
\right)
+ C_{\rho\alpha\sigma\beta}.
\end{multline}
Substituting this in \Eq{eq:mcG} leads to
\begin{multline}
\oper{\mc{G}}_{\alpha\beta\gamma\delta} h^{\gamma\delta} =
\frac{1}{4}\bigg(
- \frac{1}{3}\, G h_{\alpha\beta}
- g_{\alpha\beta} G_{\rho\sigma} h^{\rho\sigma}
+ \frac{1}{3}\, g_{\alpha\beta} G h
\\+ G_{\alpha\rho} {h_\beta}^\rho
+ G_{\rho\beta} {h^\rho}_\alpha
+ 2 C_{\rho\alpha\sigma\beta} h^{\rho\sigma} \bigg).
\label{eq:mcG1}
\end{multline}

\subsection{Wave equation}
\label{sec:weq}

Using the above notation, the general wave equation~\eq{eq:DX} can be written as follows:
\begin{gather}
\oper{D}_{\alpha\beta\gamma\delta} h^{\gamma\delta} = 0,
\label{eq:euler0}
\end{gather}
where we have introduced
\begin{gather}
\oper{D}_{\alpha\beta\gamma\delta} \doteq (\oper{D}^{\rm vac} + \oper{\mc{G}} + \oper{D}^{\rm m})_{\alpha\beta\gamma\delta}.
\end{gather}
Then, two distinct regimes are possible depending on the magnitude of $C_{\alpha\beta\gamma\delta}$ relative to $R_{\alpha\beta} \sim G_{\alpha\beta} = \mc{O}(\rho) \sim \oper{D}^{\rm m}_{\alpha\beta\gamma\delta}$. If $C_{\alpha\beta\gamma\delta} \gg R_{\alpha\beta}$, then \(\oper{\mc{G}}_{\alpha\beta\gamma\delta}\) is dominated by the Weyl tensor and $\oper{\mc{G}}_{\alpha\beta\gamma\delta} \gg \oper{D}^{\rm m}_{\alpha\beta\gamma\delta}$, so the interaction with matter is insignificant. If $C_{\alpha\beta\gamma\delta}\lesssim R_{\alpha\beta}$, then $R_{\alpha\beta\gamma\delta} \sim R_{\alpha\beta}$, so $\oper{\mc{G}}_{\alpha\beta\gamma\delta} = \mc{O}(\rho)$. Because \Eq{eq:gder} implies $R_{\alpha\beta\gamma\delta} = \mc{O}(\ell_g^{-2})$, one also has
\begin{gather}
\ell_g^{-2} \sim \rho.
\label{eq:lgrho}
\end{gather}
It is this, second, regime that will be assumed below. 

In the presence of matter, $\oper{D}^{\rm m}_{\alpha\beta\gamma\delta}$ generally scales like~$\rho$. This is of the same order as $\oper{\mc{G}}_{\alpha\beta\gamma\delta}$, which is negligible within GO (see below). Hence, one must either give up the GO approximation or neglect the coupling with matter completely, the latter leading to exactly the same modes as in vacuum, which are briefly discussed in \Sec{sec:vacuum}. (This fact was also pointed out in \Ref{ref:asseo76}, but it is usually ignored in literature.) However, the GW--matter coupling still can be described within the GO approximation if there is an additional large dimensionless parameter that makes $\oper{D}^{\rm m}_{\alpha\beta\gamma\delta}$ much larger than $\oper{\mc{G}}$ even though both scale linearly with $\rho$. 

\section{Short-wavelength approximation}
\label{sec:go}

To simplify the general wave equation \eq{eq:euler0}, let us assume that a GW is quasimonochromatic,
\begin{gather}
h_{\alpha\beta} = \Re(\ee^{\ii\theta}a_{\alpha\beta}),
\label{eq:harmonic}
\end{gather}
where \(\theta\) is a rapid phase and $a_{\alpha\beta}$ is a slow envelope. The GW local wavevector is defined as
\begin{gather}
k_\alpha \doteq \pd_\alpha \theta = \nabla_\alpha \theta \sim \ell_h^{-1}
\end{gather}
and is assumed to change slowly on the scale comparable to that of $a_{\alpha\beta}$. Then, $\epsilon \sim \ell_h/\ell_g \ll 1$ serves as the GO parameter. Together with the assumptions \eq{eq:rhoh} and \eq{eq:lgrho}, our ordering is thereby summarized as follows:
\begin{gather}
a \ll \epsilon \sim \ell_h/\ell_g \sim \ell_h \sqrt{\rho} \ll 1.
\label{eq:scale}
\end{gather}
Then, \Eq{eq:euler0} can be written as
\begin{gather}\label{eq:euler1}
D^{(0)}_{\alpha\beta\gamma\delta} a^{\gamma\delta} + \frac{1}{4}\,M_{\alpha\beta} = 0.
\end{gather}
Here, we have introduced
\begin{multline}
D^{(0)}_{\alpha\beta\gamma\delta} \doteq \frac{1}{4} \,\big(
- k^2 g_{\alpha\gamma}g_{\beta\delta}
+ g_{\alpha\beta}g_{\gamma\delta} k^2
- k_\alpha k_\beta g_{\gamma\delta}
\\
- g_{\alpha\beta} k_\gamma k_\delta
+ k_\alpha k_\gamma g_{\beta\delta}
+ k_\beta  k_\delta g_{\alpha\gamma}
\big),
\end{multline}
$k^2 \doteq k_\mu k^\mu$, and also the GW--matter coupling term:
\begin{align}
M_{\alpha\beta}
\doteq &  \,\,
{\ampder_{\alpha\beta\rho}}^\rho
- g_{\alpha\beta} g^{\rho\sigma} {\ampder_{\rho\sigma\lambda}}^\lambda
+ g^{\rho\sigma} \ampder_{\rho\sigma\alpha\beta}
\nonumber\\ 
&+ g_{\alpha\beta} {\ampder_{\rho\sigma}}^{\rho\sigma}
- {\ampder_{\rho\beta\alpha}}^\rho
- {\ampder_{\rho\alpha\beta}}^\rho
+ 2 a^{\rho\sigma} C_{\rho\alpha\sigma\beta}
\nonumber \\ 
&
- g_{\alpha\beta} G_{\rho\sigma} a^{\rho\sigma}
+ {a_\beta}^\rho G_{\alpha\rho}
+ {a^\rho}_\alpha G_{\rho\beta}
\notag \\
&
-(a_{\alpha\beta}- a g_{\alpha\beta}) G/3
+ 4 \ee^{-\ii\theta} \oper{D}^{\rm m}_{\alpha\beta\gamma\delta} (\ee^{\ii\theta}a^{\gamma\delta}),
\label{eq:Mdef}
\end{align}
where \m{\ampder_{\alpha\beta\mu\nu} \doteq
\pd_\nu \pd_\mu a_{\alpha\beta}
+ 2 \ii k_{(\mu} \pd_{\nu)} a_{\alpha\beta}
+ \ii a_{\alpha\beta} \pd_\mu k_\nu
}; \ie
\begin{gather}\label{eq:Mdef0}
M_{\alpha\beta} = 4 D^{\text{m}}_{\alpha\beta\gamma\delta} a^{\gamma\delta} + \mc{O}(\epsilon a),
\end{gather}
where \m{D^{\text{m}}_{\alpha\beta\gamma\delta}} is the Weyl symbol of \m{\oper{D}^{\text{m}}_{\alpha\beta\gamma\delta}} \cite{my:quasiop1}. Under the assumed ordering, the second term in \Eq{eq:Mdef0} is small compared with \m{D^{(0)}_{\alpha\beta\gamma\delta}a^{\gamma\delta}}. Even when this term is neglected, though, \Eq{eq:euler1} may not be easy to solve, because \m{D^{\text{m}}_{\alpha\beta\gamma\delta}} does not have an obvious structure in the general case. Much like in plasma-wave theory \cite{book:stix}, symmetry considerations are, in general, not enough to find the wave polarization. Also note that the term \m{\mc{O}(\epsilon a)} in \Eq{eq:Mdef0} may not be small compared with \m{D^{\text{m}}_{\alpha\beta\gamma\delta} a^{\gamma\delta}}, so let us retain it for~now.

Let us proceed as done for vacuum waves in \Ref{ref:maccallum73}, which in turn follows the methodology from \Ref{ref:choquet69}. Consider the trace-reverse of \Eq{eq:euler1},
\begin{gather}
k_\alpha k^\rho a_{\rho\beta}
+ k_\beta k^\rho a_{\rho\alpha}
- k_\alpha k_\beta a
- k^2 a_{\alpha\beta} 
+ \bar{M}_{\alpha\beta}
= 0.
\end{gather}
Using the trace-reversed amplitude \m{\bar{a}_{\alpha\beta}}, this can also be represented as
\begin{gather}
k_\alpha k^\rho \bar{a}_{\rho\beta}
+ k_\beta k^\rho \bar{a}_{\rho\alpha}
- k^2 \bar{a}_{\alpha\beta}
+ \frac{g_{\alpha\beta}}{2}\, k^2 \bar{a} 
+ \bar{M}_{\alpha\beta}
= 0,
\label{eq:euler2}
\end{gather}
or equivalently,
\begin{gather}
k_\alpha k^\rho \bar{a}_{\rho\beta}
+ k_\beta k^\rho \bar{a}_{\rho\alpha}
- k^2 a_{\alpha\beta}
+ \bar{M}_{\alpha\beta}
= 0,
\label{eq:euler3}
\end{gather}
the contraction of which gives
\begin{gather}
k^\alpha k^\beta \bar{a}_{\alpha\beta}
+ \frac{1}{2}\, k^2 \bar{a}
= \frac{M}{2}.
\label{eq:eulertrace}
\end{gather}
For GWs that interact with matter, $k^2$ is nonzero, so one can introduce the projection tensor
\begin{gather}
\Pi^{\alpha\beta} \doteq
g^{\alpha\beta} - \frac{k^\alpha k^\beta}{k^2}.
\label{eq:Pi}
\end{gather}
(Vacuum waves can be considered as a limit $k^2 \to 0$; cf.\ \Ref{tex:mydecomp}.) A straightforward calculation shows that
\begin{multline}
k^2 \Pi^{\rho\alpha} \Pi^{\sigma\beta} \bar{a}_{\rho\sigma}
= -k^\alpha k_\rho \bar{a}^{\rho\beta}
- k^\beta k_\rho \bar{a}^{\rho\alpha}
\\
+ k^2 \bar{a}^{\alpha\beta}
+ \bar{a}_{\rho\sigma}k^\rho k^\sigma
\frac{k^\alpha k^\beta}{k^2}.
\end{multline}
Then, also using \Eq{eq:eulertrace}, one can rewrite \Eq{eq:euler2}~as
\begin{multline}
- k^2 \Pi^{\rho\alpha} \Pi^{\sigma\beta} \bar{a}_{\rho\sigma} 
+ \frac{1}{2}\, k^2 g^{\alpha\beta}\bar{a} 
- \frac{\bar{a}}{2}\, k^\alpha k^\beta
\\
+ \frac{M}{2 k^2}\, k^\alpha k^\beta
+ \bar{M}^{\alpha\beta}
= 0,
\end{multline}
or more succinctly as,
\begin{gather}
- k^2 \Pi^{\rho\alpha} \Pi^{\sigma\beta} \bar{a}_{\rho\sigma}
+ \frac{1}{2}\, k^2 \Pi^{\alpha\beta} \bar{a} 
-\frac{1}{2}\, \Pi^{\alpha\beta}M
= -M^{\alpha\beta}.
\end{gather}
The above equation can be further rewritten as
\begin{gather}
\Pi^{\rho\alpha} \Pi^{\sigma\beta}
\left(
k^2 a_{\rho\sigma}
+\frac{1}{2}\, M g_{\rho\sigma}
\right)
= M^{\alpha\beta},
\label{eq:euler4}
\end{gather}
where we used
\begin{gather}
\Pi^{\alpha\rho} {\Pi_\rho}^\beta
\equiv \Pi^{\alpha\rho}\Pi^{\sigma\beta}g_{\rho\sigma} = \Pi^{\alpha\beta}.
\label{eq:pipi}
\end{gather}
Because \Eq{eq:euler4} is symmetric with respect to interchanging $\alpha \leftrightarrow \beta$, it represents a total of ten equations. Also note that, \(M^{\alpha\beta}\) can be decomposed as
\begin{multline}
M^{\alpha\beta}
= \Pi^{\rho\alpha} \Pi^{\sigma\beta} M_{\rho\sigma}
+ \frac{2}{k^2}\,k_\rho M^{\rho(\alpha} k^{\beta)} 
\\
- \frac{1}{k^4}\,k_\rho M^{\rho\sigma} k_\sigma k^\alpha k^\beta
\end{multline}
[as proven by direct substitution of \Eq{eq:Pi}], which can be used to write \Eq{eq:euler4} as
\begin{multline}
\Pi^{\alpha\rho} \Pi^{\beta\sigma}
\left(
k^2 a_{\rho\sigma}
- \bar{M}_{\rho\sigma}
\right)
= \frac{2}{k^2}\,k_\rho M^{\rho(\alpha} k^{\beta)} 
\\
- \frac{1}{k^4}\,k_\rho M^{\rho\sigma} k_\sigma k^\alpha k^\beta.
\label{eq:euler5}
\end{multline}

Equation \eq{eq:euler5} can be decomposed into a \textit{longitudinal} part and a \textit{transverse} part defined as follows. The longitudinal part can be obtained by multiplying \Eq{eq:euler5} with \(k_\alpha\). The left-hand side vanishes then, and the right-hand side yields
\begin{gather}
k_\alpha M^{\alpha\beta} = 0.
\label{eq:eulerlong}
\end{gather}
The transverse part of the wave equation can be obtained by multiplying \Eq{eq:euler5} with \(\Pi^\gamma{}_\alpha \Pi_\beta{}^\delta\), which eliminates the right-hand side, yielding
\begin{gather}
\Pi^{\gamma\rho} \Pi^{\delta\sigma}
\left(
k^2 a_{\rho\sigma}
- \bar{M}_{\rho\sigma}
\right)
= 0,
\label{eq:eulertrans0}
\end{gather}
where we used \Eq{eq:pipi}. The ``general solution'' to the above equations is
\begin{gather}
k^2 a_{\rho\sigma} - \bar{M}_{\rho\sigma}[a_{\alpha\beta}] = \lambda_\rho k_\sigma + \lambda_\sigma k_\rho,
\label{eq:eulertrans1}
\end{gather}
where we used the symmetry of \(a_{\rho\sigma}\) and \(M_{\rho\sigma}\), and \(\lambda_\rho\) are constants determined by the longitudinal equations \eq{eq:eulerlong}. (The bracket $[a_{\alpha\beta}]$ has been added as a reminder that $M_{\rho\sigma}$ depends on $a_{\alpha\beta}$.) It can be easily seen, either from substitution of \Eq{eq:eulertrans1} in \Eq{eq:eulerlong} or by direct comparison of \Eq{eq:eulertrans1} with \Eq{eq:euler3}, that
\begin{gather}
\lambda_\sigma = k^\rho \bar{a}_{\rho\sigma},
\label{eq:lambda}
\end{gather}
which are the degrees of freedom always afforded by gauge invariance \cite[Sec.~8.3]{book:schutz}. Hence, \Eq{eq:eulertrans0} encodes all the physical information required to determine the solution for the perturbation, and \Eq{eq:eulerlong} serves as a check to ensure the gauge invariance of the dispersion operator.

Also notice the following. As discussed in \Sec{sec:beq}, the linear wave equation is invariant with respect to gauge transformations \eq{eq:htr}. Within the GO limit, the wave equation is \Eq{eq:euler3}, and \m{D^{(0)}_{\alpha\beta\gamma\delta}a^{\gamma\delta}} is gauge-invariant by itself, as is well known from vacuum-GW theory and also easy to check. Thus, so must be \m{\bar{M}_{\alpha\beta}}. This means that for \m{\xi^\alpha = \Re(-\ii \Lambda^\alpha \ee^{\ii\theta})}, one has
\begin{gather}
M_{\rho\sigma}[\Lambda_\alpha k_\beta + \Lambda_\beta k_\alpha] = 0,
\label{eq:Mginv}
\end{gather}
where, again, the square brackets denote the argument. This is equivalent to \Eq{eq:eulerlong} because of \Eqs{eq:Dsym}. Equations \eq{eq:eulerlong} and \eq{eq:Mginv} can be used to gauge the accuracy of approximate models of GWs, as elaborated in the following sections. 

\section{Example 1: gravitational waves in vacuum}
\label{sec:vacuum}

In a flat Minkowski space in the absence of matter, one has \m{M_{\alpha\beta} = 0}, so both \Eq{eq:eulerlong} and \Eq{eq:Mginv} are trivially satisfied. Also, \Eq{eq:euler1} becomes
\begin{gather}
k_\alpha k^\rho a_{\rho\beta}
+ k_\beta k^\rho a_{\rho\alpha}
- k_\alpha k_\beta a
- k^2 a_{\alpha\beta}
= 0.
\end{gather}
This equation was studied, for example, in \Ref{ref:maccallum73}; see also \Refs{tex:mydecomp, tex:mygwquasi}. At nonzero $k^2$, the only possible waves are coordinate waves, \ie those that can be eliminated by a coordinate transformation. At $k^2 = 0$, one finds the two usual tensor modes \cite[Eq.~(7.108)]{book:carroll} with
\begin{gather}
\omega^2 = \msf{k}^2,
\end{gather}
where we assumed the parametrization \(k_\alpha = (-\omega, 0, 0, \msf{k})\). These waves can be modified by a nonzero background Weyl tensor $C_{\alpha\beta\gamma\delta}$, which can give rise to nonzero $\oper{\mc{G}}$ (\Sec{sec:weq}). However, $C_{\alpha\beta\gamma\delta} = \mc{O}(\ell_g^{-2})$, so it is of the same order as, for example, the second-order derivatives of the envelope, which are negligible within GO. Hence, the effect of the background Weyl tensor on vacuum GWs can be described only beyond GO, \ie diffraction must be taken into account.

\section{Example 2: gravitational waves in a neutral gas}
\label{sec:examples}

\subsection{Gravitational susceptibility}
\label{sec:matter}

Now let us consider GWs in a neutral gas. For simplicity, we assume the gas to contain single species with the distribution function \(f(\vec{p})\) normalized to the local proper mass density \(\rho\) of this species. (Generalization to multiple species is straightforward.) Specifically, this means
\begin{gather}
\int \frac{\dd\vec{p}}{p^0}\,f(\vec{p}) = \frac{\rho}{m^2},
\end{gather}
where $m$ is the particle mass, and \Eq{eq:backg} yields
\begin{gather}
G_{\alpha\beta}
= T_{\alpha\beta}
=\int \mc{T}_{\alpha\beta} f(\vec{p})\,\frac{\dd\vec{p}}{p^0}.
\label{eq:linearback1}
\end{gather}
Here, the slow dependence on spacetime coordinates $x$ is assumed but not emphasized. Also, the bold font is used to denote three-dimensional (spatial) vectors, \(\dd\vec{p}/p^0\) is a Lorentz-invariant measure \cite[Eq.~(2.40)]{book:peskin}, and $p^0$ is calculated from the spatial momenta using \(p_\alpha p^\alpha = -m^2\). Since $D^{\rm m}_{\alpha\beta\gamma\delta}$ already scales linearly with \(\rho \sim \epsilon^2\), it can be calculated to the zeroth order in \m{\epsilon}, \ie as in flat spacetime. Then locally, one can adopt the Minkowski metric and
\begin{gather}
 p^0 \approx \sqrt{m^2 + \vec{p}^2}.
\end{gather}
As shown in \Ref{my:gwponder}, the corresponding \m{D^{\rm m}_{\alpha\beta\gamma\delta}} can be expressed as
\begin{gather}
D^{\rm m}_{\alpha\beta\gamma\delta}
= \int
\bigg(
\frac{\vec{k}\cdot \pd_\vec{p} f}{\omega - \vec{k} \cdot \vec{v}}\,
\mc{T}_{\alpha\beta} \mc{T}_{\gamma\delta}
+ f J_{\alpha\beta\gamma\delta} 
\bigg) \frac{\dd\vec{p}}{4(p^0)^2}
\label{eq:lm}
\end{gather}
(in a multi-species gas, summation over species should be added on the right-hand side), where \m{\mc{T}_{\alpha\beta} \doteq p_\alpha p_\beta}, \m{\vec{v} = \vec{p}/p^0}, the parametrization \(k_\alpha = (-\omega, 0, 0, \msf{k})\) is assumed again, and
\begin{gather}
J_{\alpha\beta\gamma\delta} \doteq 
\frac{\pd(\mc{T}_{\alpha\beta} \mc{T}_{\gamma\delta})}{\pd p_0}
-
\frac{g^{00}}{p^0}\,\mc{T}_{\alpha\beta} \mc{T}_{\gamma\delta}
- 4 p^0 Q_{\alpha\beta\gamma\delta},
\end{gather}
where we have introduced
\begin{gather}
Q_{\alpha\beta\gamma\delta} \doteq  \frac{g_{(\alpha\delta}\mc{T}_{\beta)\gamma}+g_{(\alpha\gamma}\mc{T}_{\beta)\delta}}{2}.
\label{eq:Q}
\end{gather}
As usual, the expression featuring the resonant denominator \m{\omega - \vec{k} \cdot \vec{v}} holds at \m{\Im\omega > 0}, and the analytic continuation of \eq{eq:lm} should be used otherwise \cite{book:stix, my:nonloc}. This means that the integration should be done over the Landau contour \m{\mc{L}}, which goes below the pole \cite{book:stix}.

To the extent that the interaction with resonant particles can be ignored, though, the integral in \eq{eq:lm} is also convergent as is. Then, one can integrate by parts and obtain (cf.\ Eq.~(120) from \Ref{my:gwponder})
\begin{align}
D^{\rm m}_{\alpha\beta\gamma\delta}
h^{\gamma\delta}
= & - \frac{1}{2} \int \frac{\dd\vec{p}}{p^0}\,f(\vec{p})\,
\Big[
\mc{T}_{\beta\rho} {h^\rho}_{\alpha}
+\mc{T}_{\alpha\rho} {h_{\beta}}^\rho
\notag\\
& + \frac{k^2}{2\Omega^2}\,\mc{T}_{\alpha\beta} \mc{T}_{\rho\sigma} h^{\rho\sigma} 
-\frac{1}{\Omega}\,\mc{T}_{\alpha\beta} \Omega_{\rho\sigma} h^{\rho\sigma}
\notag\\
& -\frac{1}{\Omega}\,\Omega_{\alpha\beta} \mc{T}_{\rho\sigma} h^{\rho\sigma}
\Big],
\label{eq:Dm2}
\end{align}
where we introduced
\begin{gather}
\Omega_{\alpha\beta} \doteq p_{(\alpha} k_{\beta)},
\quad
\Omega \doteq g^{\alpha\beta} \Omega_{\alpha\beta}.
\end{gather}
%

\subsection{Gravitostatic modes}
\label{sec:jeans}

As readily seen from \Eq{eq:Dm2}, the matrix \m{D^{\rm m}_{\alpha\beta\gamma\delta}} is of order \m{\rho N^2}, where $N \doteq \msf{k}/\omega$ is the refractive index. Then, as follows from our earlier argument (\Sec{sec:weq}), coupling with matter can be described within the GO approximation if
\begin{gather}\label{eq:N}
N \gg 1.
\end{gather}
To assess the gauge invariance of this reduced theory, let us consider an alternative representation of \Eq{eq:lm} as described in \cite[Eqs.~(117)--(120)]{my:gwponder}, the derivation of which can be found in \cite[Appendix~B]{my:gwponder}:
\begin{gather}
D^{\rm m}_{\alpha\beta\gamma\delta}
= \frac{1}{4}\int
\frac{\dd\vec{p}}{p^0}\,f(\vec{p})\,
\bigg[
k_\rho \frac{\pd}{\pd p_\rho} \left(\frac{\mc{T}_{\alpha\beta} \mc{T}_{\gamma\delta}}{k_\sigma p^\sigma} \right)
- 4 Q_{\alpha\beta\gamma\delta}
\bigg].
\label{eq:lm1}
\end{gather}
In the limit \eq{eq:N}, \Eq{eq:Mdef0} can be written as
\begin{gather}\label{eq:Mappr}
M_{\alpha\beta} = 4D^{\text{m}}_{\alpha\beta\gamma\delta} a^{\gamma\delta}
+ \delta M_{\alpha\beta}
\end{gather}
and \m{D^{\rm m}_{\alpha\beta\gamma\delta}} can be approximated as
\begin{gather}\label{eq:Dappr0}
D^{\rm m}_{\alpha\beta\gamma\delta}
\approx \frac{1}{4}\int
\frac{\dd\vec{p}}{p^0}\,f(\vec{p})\,
\left[
k_\rho \frac{\pd}{\pd p_\rho} \left(\frac{\mc{T}_{\alpha\beta} \mc{T}_{\gamma\delta}}{k_\sigma p^\sigma} \right)
\right],
\end{gather}
where the subdominant terms from \Eqs{eq:Mdef0} and \eq{eq:lm1} are subsumed under \(\delta M_{\alpha\beta} = \mc{O}(N^0)\), which can be neglected, and the dominant term is \m{4 D^{\text{m}}_{\alpha\beta\gamma\delta} a^{\gamma\delta} = \mc{O}(N^2)}.
It is readily seen then that
\begin{align}\label{eq:check}
D^{\rm m}_{\alpha\beta\gamma\delta}k^\alpha
&& = \frac{k^\alpha}{4}\int
\frac{\dd\vec{p}}{p^0}\,f(\vec{p})\,
\left[
k_\rho \frac{\pd}{\pd p_\rho} \left(\frac{\mc{T}_{\alpha\beta} \mc{T}_{\gamma\delta}}{k_\sigma p^\sigma} \right)
\right]
\nonumber \\ &&
= \frac{1}{4}\int
\frac{\dd\vec{p}}{p^0}\,f(\vec{p})\,
\left[
k_\rho \frac{\pd \left(p_\beta p_\gamma  p_\delta \right)}{\pd p_\rho}
\right]
\nonumber \\ &&
= \frac{1}{4} \left(
k_\beta T_{\gamma\delta}
+ k_\gamma T_{\beta\delta}
+ k_\delta T_{\beta\gamma}
\right),
\end{align}
which, like \m{\delta M_{\alpha\beta}k^\alpha}, is \m{\mc{O}(N^0 \msf{k})}. Hence, \m{D^{\rm m}_{\alpha\beta\gamma\delta}k^\alpha} can be ignored up to \m{\delta M_{\alpha\beta}k^\alpha} which is negligible under the assumption \eq{eq:N}. Because the approximate \m{D^{\rm m}_{\alpha\beta\gamma\delta}} \eq{eq:lm1} is symmetric in its four indices, both \Eq{eq:eulerlong} and \Eq{eq:Mginv} are satisfied. This makes \Eq{eq:Dappr0} a satisfactory approximation. By analogy with electrostatic waves in plasmas, GWs that satisfy this approximation can be called \textit{gravitostatic} (and the Newtonian limit corresponds to $N \to \infty$). Similarly, \m{D^{\rm m}_{\alpha\beta\gamma\delta}} can be also approximated~as
\begin{gather}
D^{\rm m}_{\alpha\beta\gamma\delta}
= \int
\frac{\vec{k}\cdot \pd_\vec{p} f}{\omega - \vec{k} \cdot \vec{v}}\,
\mc{T}_{\alpha\beta} \mc{T}_{\gamma\delta}
\,\frac{\dd\vec{p}}{4(p^0)^2},
\label{eq:Dappr1}
\end{gather}
which is equal to \Eq{eq:Dappr0} up to subdominant terms of \(\mc{O}(N^0)\) as can be seen by comparing \Eqs{eq:lm} and \eq{eq:lm1}.

One can expect gravitostatic modes to be the adiabatic modes of the Newtonian Jeans theory in the absence of the Hubble expansion \cite[Sec.~6.2.1]{book:mukhanov}. These modes are derived from our general formulation as follows. Under the Newtonian limit, the background energy tensor can be considered to be nonrelativistic:
\begin{gather}
\frac{\mc{T}_{\alpha\beta}\mc{T}_{\gamma\delta}}{(p^0)^2} \approx m^2 \delta^0_\alpha \delta^0_\beta \delta^0_\gamma \delta^0_\delta.
\label{eq:nonrel}
\end{gather}
Let us also change the normalization of the distribution function as \(f(\vec{p})\,\dd\vec{p} \to \rho(x) m^{-1} f(\vec{v})\,\dd\vec{v}\). Then, \Eqs{eq:Dappr1} and \eq{eq:Mappr} become
\begin{gather}
D^{\rm m}_{\alpha\beta\gamma\delta}
\approx 
\mc{X} \delta^0_\alpha \delta^0_\beta \delta^0_\gamma \delta^0_\delta,
\quad
M_{\alpha\beta} = 4 \mc{X} \delta^0_\alpha \delta^0_\beta a^{00},
\label{eq:mquasi}
\end{gather}
where we have introduced
\begin{gather}
\mc{X}
\doteq \frac{\rho}{4} \int_{\mc{L}}
\frac{\vec{k}\cdot \pd_\vec{v} f(\vec{v})}{\omega - \vec{k} \cdot \vec{v}}\,
\dd \vec{v}.
\label{eq:X}
\end{gather}

Because \Eqs{eq:Mginv} and \eq{eq:eulerlong} are satisfied, one can adopt any gauge. We choose \(\lambda_\beta = 0\) in \Eq{eq:lambda}, which corresponds to the Lorenz gauge:
\begin{gather}\label{eq:lgauge}
k^\alpha \bar{a}_{\alpha\beta} = 0.
\end{gather}
From \Eq{eq:mquasi}, one finds that 
\begin{gather}
\bar{M}_{\alpha\beta}
= 2 \mc{X} \left(2 \delta^0_\alpha \delta^0_\beta
+ \eta_{\alpha\beta}
\right) a^{00} 
= 2\mc{X} \I_{\alpha\beta} a_{00},
\end{gather}
where \(\I_{\alpha\beta}\) is the identity matrix. Using this and \Eq{eq:lgauge}, one obtains from \Eq{eq:euler3} that
\begin{gather}\label{eq:jpol}
k^2 a_{\alpha\beta} = 2\mc{X} \I_{\alpha\beta} a_{00}.
\end{gather}
This means that gravitostatic waves have a longitudinal polarization, namely,
\begin{gather}
a_{\alpha\beta} = \text{diag}\,\lbrace 1, 1, 1, 1\rbrace \times \text{const}.
\label{eq:jeanspol}
\end{gather}
Also, substituting \m{\alpha = \beta = 0} into \Eq{eq:jpol}, one finds that
\begin{gather}
2\mc{X} = k^2 \approx \msf{k}^2,
\end{gather}
where the approximate equality is due to \Eq{eq:N}. Finally, using \Eq{eq:X}, one obtains the following dispersion relation:
\begin{gather}
1 - \frac{\rho}{2 \msf{k}^2} \int_{\mc{L}} \dd \vec{v}\,
\frac{\vec{k}\cdot \pd_\vec{v} f(\vec{v})}{\omega - \vec{k} \cdot \vec{v}}
\approx 0,
\label{eq:jeanskin}
\end{gather}
One can recognize \Eq{eq:jeanskin} as the dispersion relation of the kinetic Jeans mode \cite{ref:lima02, ref:trigger04, ref:ershkovich08}.

Equation \eq{eq:jeanskin} is identical to the dispersion relation of Langmuir oscillations in nonrelativistic collisionless plasma with plasma frequency $\omega_p$ up to replacing $\omega_p^2$ with $- \omega_J^2$, where
\begin{gather}
\omega_J \doteq \sqrt{\frac{\rho}{2}}
\end{gather}
(or \(\omega_J = \sqrt{4 \pi \gc \rho}\), in units when the gravitational constant \(\gc\) is not equal to one) is the Jeans frequency. Hence, the limiting cases of \Eq{eq:jeanskin} are readily obtained in the same way \cite{book:stix}. In cold gas, where the typical velocities satisfy $v \ll \omega/\msf{k}$, one can use
\begin{align}
\int_{\mc{L}} \dd \vec{v}\,
\frac{\vec{k}\cdot \pd_\vec{v} f(\vec{v})}{\omega - \vec{k} \cdot \vec{v}}
& \approx \int \dd \vec{v}\left(1 + \frac{\vec{k}\cdot \vec{v}}{\omega}\right)
\frac{\vec{k}}{\omega}\cdot \pd_\vec{v} f(\vec{v})
\nonumber\\
& = \frac{1}{\omega^2} \int \dd \vec{v}\,
\left( \vec{k}\cdot \vec{v}\right) \vec{k}\cdot \pd_\vec{v} f(\vec{v})
\nonumber\\
& = -\frac{\msf{k}^2}{\omega^2} \int \dd \vec{v}\,f(\vec{v})
\nonumber\\
& = -\frac{\msf{k}^2}{\omega^2},
\label{eq:N2}
\end{align}
so \Eq{eq:jeanskin} leads to the well-known formula \cite{ref:thompson08}
\begin{gather}
\omega^2 \approx -\omega_J^2.
\end{gather}
[Also note that $N^2 =  2\msf{k}^2/\rho$, so \Eq{eq:N} is always satisfied provided that $\msf{k}^2$ is nonzero and $\rho$ is small enough.] By keeping terms $\mc{O}[(\msf{k} v_T/\omega_J)^3]$ in the above expansion, one can also obtain a more general result (cf.\ \Ref{ref:bohm49}):
\begin{gather}\label{eq:omfl}
\omega^2 \approx -\omega_J^2 + 3 \msf{k}^2 v_T^2,
\quad 
\textstyle
v_T^2 \doteq \int \dd v\, v_z^2 f(\vec{v}).
\end{gather}

\begin{figure}
\includegraphics[width = .46\textwidth]{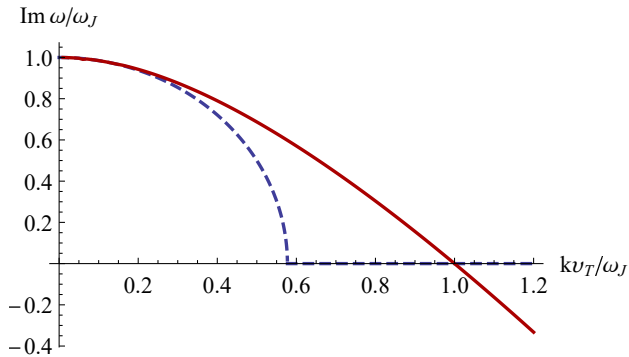}
\caption{Solid red -- a numerical solution of \Eq{eq:jeansM} for the Jeans-instability growth rate $\Im \omega$ in units $\omega_J$ vs.~$\msf{k} v_T/\omega_J$. Dashed blue -- a solution of \Eq{eq:omfl}, which is the correct asymptotic approximation of the exact solution at small~$\msf{k} v_T/\omega$.}
\label{fig:jeans}
\end{figure}

Let us also consider the case of an isotropic Maxwellian distribution
\begin{gather}
f(\vec{v}) = \frac{1}{(2\pi v_T)^{3/2}}\,\exp\left(-\frac{v^2}{2v_T^2}\right).
\end{gather}
Using the aforementioned analogy between the Jeans mode and Langmuir oscillations, one can readily express \Eq{eq:jeanskin} through the plasma dispersion function \cite{book:stix}
\begin{gather}
Z_0(\zeta) \doteq \ii\,\text{sgn}(\msf{k})\sqrt{\pi}\,\ee^{-\zeta^2} - 2\mc{S}(\zeta),
\\
\textstyle
\mc{S}(\zeta) \doteq \ee^{-\zeta^2} \int_0^\zeta \dd z\,\ee^{z^2}.
\end{gather}
where \m{\mc{S}} is known as the Dawson function. Specifically, \Eq{eq:jeanskin} becomes
\begin{gather}\label{eq:jeansM}
1 + \frac{\omega_J^2}{2\msf{k}^2 v_T^2}\,\frac{\dd Z_0(\zeta)}{\dd \zeta} \approx 0,
\quad
\zeta \doteq \frac{\omega}{\msf{k} v_T \sqrt{2}}.
\end{gather}
A numerical solution of this equation is shown in \Fig{fig:jeans} (cf.\ the qualitative figure in \Ref{ref:trigger04}). Unlike within the fluid approximation \eq{eq:omfl}, \m{\Im\omega} is nonzero at all \m{|\msf{k}| \ne \omega_J/v_T}, and waves damp (\m{\Im\omega < 0}) at \m{|\msf{k}| > \omega_J/v_T}.

The agreement of the Newtonian gauge derived as usual \cite{book:carroll} and the polarization obtained here \eq{eq:jeanspol} warrants a comment. Usually, the background is fixed to be Minkowski and the curvature produced by matter is ascribed to the perturbation, leading to the polarization of \(\text{diag}\,\lbrace 1, 1, 1, 1\rbrace\) in the Newtonian limit. (Also, see the related discussion in \Sec{sec:jeansmetric}.) Hence, that polarization is due to a specific kind of source of the GWs. In our analysis, though, the slow modification of the metric produced by matter is ascribed to the background metric \m{g_{\alpha\beta}}, while the aforementioned polarization \eq{eq:jeanspol} is the property of the perturbation. 

\section{Conclusions}
\label{sec:conc}

In summary, we study the dispersion of linear GWs propagating through matter. Our model accounts both for metric oscillations and the backreaction of matter on these oscillations, so the usual tensor modes and the gravitational modes strongly coupled with matter are treated on the same footing. Using the averaged-Lagrangian approach, the GW equation \eq{eq:euler0} [see also \Eqs{eq:dm0}, \eq{eq:dvac} and \eq{eq:mcG1}] is derived, which also accounts for the effect of the background-metric inhomogeneity, including the Weyl curvature. A test [\Eqs{eq:eulerlong} and \eq{eq:Mginv}] is proposed for accessing the gauge invariance of models of the matter polarizability. Next, the wave equation is studied within the short-wavelength limit \eq{eq:euler1}. We show that the effect of matter on the tensor modes is comparable to diffraction and therefore negligible within the GO approximation. However, this approximation is applicable to modes strongly coupled with matter due to their large refractive index \(N\). (By analogy with electrostatic waves in plasmas, GWs in this limit can be called gravitostatic, with the Newtonian limit corresponding to \m{N \to \infty}.) GWs in ideal gas are studied using the corresponding gravitational polarizability \eq{eq:lm}, which we derived earlier in \Ref{my:gwponder}. This formulation subsumes the Jeans instability (\Sec{sec:jeans}) as a collective GW mode with a peculiar polarization \eq{eq:jeanspol}, which is derived from the dispersion matrix rather than assumed a~priori. This forms a foundation for systematically extending GW theory to GW interactions with plasmas, where symmetry considerations alone are insufficient to predict the wave polarization \cite{book:stix}.

This material is based upon the work supported by National Science Foundation under the grant No. PHY~1903130.

\appendix

\section{Second-order Einstein--Hilbert action}
\label{app:ricciscalar}

The existing derivations of the second-order Einstein--Hilbert action \m{S^{(2)}_{\rm EH}} \eq{eq:2leh} are typically restricted to vacuum settings, omit significant details, or do not pay enough attention to the numerical coefficients that are important for studying the GW--matter coupling, as elaborated in \Sec{sec:S2}. A more comprehensive derivation is needed for our purposes and is presented below. Let us begin by considering the Lagrangian density
\begin{gather}\label{eq:LEH}
\mc{L}_{\rm EH} = \frac{\sqrt{-\total{g}}}{2}\, \total{R}
\end{gather}
that determines the full Einstein--Hilbert action \eq{eq:SEH}. The total Ricci scalar \m{\total{R}} that enters \Eq{eq:LEH} can be calculated as \m{\total{R} = \total{g}^{\alpha\beta}{\total{R}^\rho}_{\alpha\rho\beta}}, where \m{{\total{R}^\rho}_{\alpha\sigma\beta}} is the Riemann tensor associated with the full metric. This tensor can be expressed through the corresponding Christoffel symbols \m{\total{\Gamma}^\rho_{\alpha\beta}} as
\begin{gather}
{\total{R}^\rho}_{\alpha\sigma\beta}
= \pd_\sigma \total{\Gamma}^\rho_{\beta\alpha}
- \pd_\beta \total{\Gamma}^\rho_{\sigma\alpha}
+ \total{\Gamma}^\rho_{\sigma\lambda} \total{\Gamma}^\lambda_{\beta\alpha}
- \total{\Gamma}^\rho_{\beta\lambda} \total{\Gamma}^\lambda_{\sigma\alpha}.
\label{eq:Ra}
\end{gather}
Let us decompose \m{\total{\Gamma}^\rho_{\alpha\beta}} as follows:
\begin{gather}
\total{\Gamma}^\rho_{\alpha\beta}
= \Gamma^\rho_{\alpha\beta} + \widetilde{\Gamma}^\rho_{\alpha\beta},
\label{eq:Ga}
\end{gather}
where \m{\Gamma^\rho_{\alpha\beta}} are the Christoffel symbols associated with the background metric [\Eq{eq:christoffel}] and \m{\widetilde{\Gamma}^\rho_{\alpha\beta}} is the remaining perturbation, which is a proper tensor because it equals the difference of two connections. Using \Eq{eq:Ga}, one can rewrite \Eq{eq:Ra} as
\begin{multline}
{\total{R}^\rho}_{\alpha\sigma\beta} = {R^\rho}_{\alpha\sigma\beta}
+ 2\Big(
\pd_{[\sigma} \widetilde{\Gamma}^\rho_{\beta]\alpha}
+ \Gamma^\rho_{[\sigma\lambda} \widetilde{\Gamma}^\lambda_{\beta]\alpha}
\\
+ \widetilde{\Gamma}^\rho_{[\sigma\lambda} \Gamma^\lambda_{\beta]\alpha}
+ \widetilde{\Gamma}^\rho_{[\sigma\lambda} \widetilde{\Gamma}^\lambda_{\beta]\alpha}
\Big),
\label{eq:tR}
\end{multline}
where \({R^\rho}_{\sigma\alpha\beta}\) is given by \Eq{eq:riemann} and represents the Riemann tensor associated with the background metric. Equation \eq{eq:tR} can also be written as
\begin{gather}
{\total{R}^\rho}_{\alpha\sigma\beta} = {R^\rho}_{\alpha\sigma\beta}
+ 2\left(
\boldsymbol{\del}_{[\sigma} \widetilde{\Gamma}^\rho_{\beta]\alpha}
+ \widetilde{\Gamma}^\lambda_{[\sigma\alpha} \widetilde{\Gamma}^\rho_{\beta]\lambda}
\right),
\label{eq:riemanntot}
\end{gather}
where \(\boldsymbol{\del}\) denotes the covariant derivative associated with \m{\total{\Gamma}^\rho_{\alpha\beta}} and we used that \m{\total{\Gamma}^\rho_{\alpha\beta}} is torsion-free. Then, \Eq{eq:LEH} can be written as
\begin{multline}
\mc{L}_{\rm EH}
= \frac{\sqrt{-\total{g}}}{2}\, \total{g}^{\alpha\beta} R_{\alpha\beta}
+ \sqrt{-\total{g}}\, \total{g}^{\alpha\beta} \widetilde{\Gamma}^\lambda_{[\rho\alpha} \widetilde{\Gamma}^\rho_{\beta]\lambda}
\\
+ \frac{\sqrt{-\total{g}}}{2}\,\boldsymbol{\del}_{\rho}
\left(
\widetilde{\Gamma}^\rho_{\beta\alpha}
\total{g}^{\alpha\beta}
\right)
- \frac{\sqrt{-\total{g}}}{2}\,
\boldsymbol{\del}_{\beta}
\left(
\widetilde{\Gamma}^\rho_{\rho\alpha}
\total{g}^{\alpha\beta}
\right),
\end{multline}
where \m{R_{\alpha\beta} \doteq {R^\rho}_{\alpha\rho\beta}}. The last two terms contribute only boundary terms to the action \eq{eq:SEH}, so they can be ignored. Hence, one obtains \m{\mc{L}_{\rm EH} = \mc{L}_G + \mc{L}_{\rm vac}}, where
\begin{subequations}
\begin{gather}
\mc{L}_G
\doteq \frac{\sqrt{-\total{g}}}{2}\, \total{g}^{\alpha\beta} R_{\alpha\beta},
\label{eq:LGtot}
\\
\mc{L}_{\rm vac}
\doteq \sqrt{-\total{g}}\, \total{g}^{\alpha\beta} \widetilde{\Gamma}^\lambda_{[\rho\alpha} \widetilde{\Gamma}^\rho_{\beta]\lambda}.
\label{eq:Lvactot}
\end{gather}
\end{subequations}

The determinant of the full metric can be represented as \cite[Eq.~(105.4)]{book:landau2}
\begin{gather}
\total{g}
= g \left(
1 + h + \frac{h^2}{2}
- \frac{1}{2}\, h^{\alpha\beta} h_{\alpha\beta}
\right)
+ \mc{O}(a^3),
\label{eq:detg}
\end{gather}
and the inverse full metric can be expanded using \Eq{eq:hupper}. Substituting these into \Eq{eq:LGtot} leads~to
\begin{gather}
\mc{L}_G
= \mc{L}_G^{(0)}
+ \mc{L}_G^{(1)}
+ \mc{L}_G^{(2)} + \mc{O}(a^3).
\end{gather}
Here, \m{\mc{L}_G^{(0)} = \sqrt{-g}R/2} is the zeroth-order term. The next, first-order, term 
\begin{gather}
\mc{L}_G^{(1)} = \frac{\sqrt{-g}}{2}
\left(\frac{1}{2}\,R h - h^{\alpha\beta} R_{\alpha\beta}\right)
\end{gather}
does not contribute to the action integral due to \Eq{eq:havr}, so it can be ignored. The term \m{\mc{O}(a^3)} is ignorable within the accuracy of our model as well. The remaining term \m{\mc{L}_G^{(2)} = \mc{O}(a^2)} is given~by
\begin{multline}
\mc{L}_G^{(2)}
= \frac{\sqrt{-g}}{4}\bigg(
- \frac{1}{2}\, R h^{\alpha\beta} h_{\alpha\beta}
- R_{\alpha\beta} h^{\alpha\beta} h
\\
+ \frac{1}{4}\, R h^2
+ 2 R_{\alpha\beta} h^{\alpha\rho} {h_\rho}^\beta
\bigg).
\end{multline}
Rewriting it through \m{G_{\alpha\beta}}, which is given by \Eq{eq:Gabbg}, leads to \Eq{eq:LG}.

Now let us consider \m{\mc{L}_{\rm vac}} given by \Eq{eq:Lvactot}. Because this term is quadratic in \m{\widetilde{\Gamma}^\rho_{\alpha\beta}}, the leading-order approximation for the latter is sufficient and the full metric can be replaced with the background metric, so
\begin{gather}
\mc{L}_{\rm vac} \approx \mc{L}_{\rm vac}^{(2)}
\doteq \sqrt{-g} g^{\alpha\beta} \widetilde{\Gamma}^\lambda_{[\rho\alpha} \widetilde{\Gamma}^\rho_{\beta]\lambda}.
\label{eq:Lvac1}
\end{gather}
The leading-order term of \m{\widetilde{\Gamma}^\rho_{\alpha\beta}} can be calculated to yield [cf.\ \Eq{eq:christoffel}]
\begin{align}
\widetilde{\Gamma}^\rho_{\alpha\beta}
& \approx -\frac{h^{\rho\sigma}}{2}
\left(
\pd_\alpha g_{\beta\sigma} + \pd_\beta g_{\alpha\sigma} - \pd_\sigma g_{\alpha\beta}
\right)
\nonumber \\& \qquad \quad
+ \frac{g^{\rho\sigma}}{2}
\left(
\pd_\alpha h_{\beta\sigma} + \pd_\beta h_{\alpha\sigma} - \pd_\sigma h_{\alpha\beta}
\right),
\label{eq:auxG}
\end{align}
or equivalently,
\begin{gather}
\widetilde{\Gamma}^\rho_{\alpha\beta} = \frac{g^{\rho\sigma}}{2}
\left(
\pd_\alpha h_{\beta\sigma} + \pd_\beta h_{\alpha\sigma} - \pd_\sigma h_{\alpha\beta}
\right)
- h^{\rho\sigma} g_{\lambda\sigma} \Gamma^\lambda_{\alpha\beta}.
\notag
\end{gather}
Using again that the background connection is torsion-free, this can also be written as
\begin{gather}
\widetilde{\Gamma}^\rho_{\alpha\beta}
\approx \frac{g^{\rho\sigma}}{2}
\left(
\del_\alpha h_{\beta\sigma} + \del_\beta h_{\alpha\sigma} - \del_\sigma h_{\alpha\beta}
\right),
\label{eq:christperturb1}
\end{gather}
where \(\del\) denotes the covariant derivative with respect to the background connection. Then, a straightforward calculation shows that \Eq{eq:Lvac1} leads to \Eq{eq:Lvac}. 

Combined together, the above results yield that \m{S_{\rm EH} \approx S_{\rm EH}^{(0)} + S_{\rm EH}^{(2)}}, where \m{S_{\rm EH}^{(0)} = \int \dd^4 x\,\mc{L}_G^{(0)}} is independent of \m{h^{\alpha\beta}} and \m{S_{\rm EH}^{(2)}} is given by \Eq{eq:2leh}, with \m{\mc{L}_{\rm vac}^{(2)}} given by \Eq{eq:Lvac} and \m{\mc{L}_G^{(2)}} given by \Eq{eq:LG}.

\section{Alternative derivation of \Eqs{eq:dvac} and \eq{eq:mcG}}
\label{app:riccitensor}

Here, we present an alternative derivation of the wave equation in normal coordinates, characterized by \Eqs{eq:dvac} and \eq{eq:mcG}. This also serves as an independent (from \App{app:ricciscalar}) proof of \Eqs{eq:dvac} and \eq{eq:mcG}. We begin by using \Eq{eq:hupper} to write \Eq{eq:einsttensor} as
\begin{gather}
\frac{\delta \mc{L}_{\rm EH}}{\delta h^{\mu\nu}}
=\frac{\sqrt{-\total{g}}}{2} \,
\total{G}_{\alpha\beta}
\left(
-\delta^\alpha_\mu \delta^\beta_\nu
+{h_\nu}^\beta \delta^\alpha_\mu
+{h^\alpha}_\mu \delta^\beta_\nu
\right) + \mc{O}(a^2),
\label{eq:hderivative}
\end{gather}
where \m{\total{g}} is given by \Eq{eq:detg}. Note that \m{\total{G}_{\alpha\beta}} can be expanded as
\begin{multline}
\total{G}_{\alpha\beta}
= G_{\alpha\beta} + \widetilde{R}_{\alpha\beta} 
-\frac{g_{\alpha\beta}}{2} \widetilde{R}
-\frac{h_{\alpha\beta}}{2} R
\\
+\frac{g_{\alpha\beta}}{2} R_{\gamma\delta} h^{\gamma\delta}
+ \mc{O}(a^2),
\end{multline}
where \(\widetilde{R}_{\alpha\beta} \doteq {\total{R}^\rho}_{\alpha\rho\beta} - {R^\rho}_{\alpha\rho\beta}\), \(\widetilde{R} \doteq g^{\alpha\beta} \widetilde{R}_{\alpha\beta}\), and \({\total{R}^\rho}_{\alpha\sigma\beta}\) is the Riemann tensor associated with the full metric. Using this along with \Eq{eq:detg}, \Eq{eq:hderivative} can be written as
\begin{align}
\frac{\delta \mc{L}_{\rm EH}}{\delta h^{\mu\nu}}
= & \,\frac{\sqrt{-g}}{2}
\bigg(
- G_{\mu\nu}
- \widetilde{R}_{\mu\nu}
+ \frac{g_{\mu\nu}} {2}\, \widetilde{R}
+ \frac{h_{\mu\nu}}{2}\, R
\notag\\
& -\frac{g_{\mu\nu}}{2}\, h^{\gamma\delta} R_{\gamma\delta}
-\frac{h}{2}\, G_{\mu\nu}
+ G_{\mu\lambda} {h_\nu}^\lambda
+ G_{\lambda\nu} {h^\lambda}_\mu
\bigg) 
\notag\\
& + \mc{O}(a^2).
\label{eq:hderivative1}
\end{align}
Substituting \Eqs{eq:riemanntot} and \eq{eq:christperturb1} in \Eq{eq:hderivative1} and retaining only the first-order terms readily leads to \Eq{eq:Dgen}. 

One can also use normal coordinates, in which \m{\Gamma^\rho_{\alpha\beta} = 0}. Then, one obtains [cf.\ \Eq{eq:riemann}]
\begin{gather}
\widetilde{R}_{\alpha\beta}
=  \pd_\rho {\widetilde{\Gamma}^\rho_{\beta\alpha}}
- \pd_\beta {\widetilde{\Gamma}^\rho_{\rho\alpha}}.
\end{gather}
With \Eq{eq:auxG} for \m{\widetilde{\Gamma}^\rho_{\alpha\beta}}, this leads to 
\begin{align}
\widetilde{R}_{\alpha\beta}
& = \frac{g^{\gamma\delta}}{2}
\big(
\pd_\gamma \pd_\alpha h_{\beta\delta}
- \pd_\gamma \pd_\delta h_{\beta\alpha}
- \pd_\beta \pd_\alpha h_{\gamma\delta}
+ \pd_\beta \pd_\delta h_{\gamma\alpha}
\big)
\nonumber \\ 
& - \frac{h^{\gamma\delta}}{2}
\big(
\pd_\gamma \pd_\alpha g_{\beta\delta}
- \pd_\gamma \pd_\delta g_{\beta\alpha}
- \pd_\beta \pd_\alpha g_{\gamma\delta}
+ \pd_\beta \pd_\delta g_{\gamma\alpha}
\big),\notag
\end{align}
where the terms of the second and higher orders are neglected. Using \Eq{eq:gder}, one also obtains
\begin{multline}
\pd_\gamma \pd_\alpha g_{\beta\delta}
- \pd_\gamma \pd_\delta g_{\beta\alpha}
- \pd_\beta \pd_\alpha g_{\gamma\delta}
+ \pd_\beta \pd_\delta g_{\gamma\alpha}
\\
= - \frac{1}{3} \Big(
R_{\beta\alpha\delta\gamma} + R_{\beta\gamma\delta\alpha}
- R_{\beta\gamma\alpha\delta} - R_{\beta\delta\alpha\gamma}
\\
- R_{\gamma\alpha\delta\beta} - R_{\gamma\beta\delta\alpha}
+ R_{\gamma\delta\alpha\beta} + R_{\gamma\beta\alpha\delta}
\Big).
\end{multline}
Note that the above expression can be simplified considerably using the antisymmetry properties of the Riemann tensor described in, for example,  \cite[Eqs.~(3.129)--(3.132)]{book:carroll}. Then, a straightforward calculation using the same antisymmetry properties of the Riemann tensor yields
\begin{multline}
\widetilde{R}_{\alpha\beta}
= \frac{g^{\gamma\delta}}{2}
\,\big(
\pd_\gamma \pd_\alpha h_{\beta\delta}
- \pd_\gamma \pd_\delta h_{\beta\alpha}
- \pd_\beta \pd_\alpha h_{\gamma\delta}
\\
+ \pd_\beta \pd_\delta h_{\gamma\alpha}
\big)
- h^{\gamma\delta}R_{\gamma\beta\delta\alpha}.
\end{multline}
With this, \Eq{eq:hderivative1} can be readily expressed in the form
\begin{align}
\frac{\delta \mc{L}_{\rm EH}}{\delta h^{\mu\nu}}
=\frac{\sqrt{-g}}{2}
\bigg[& 
- G_{\mu\nu}
- \frac{1}{2}
\big(
\pd^\lambda \pd_\mu h_{\nu\lambda}
- \pd^\lambda \pd_\lambda h_{\nu\mu}
\notag\\
& - g^{\lambda\rho} \pd_\nu \pd_\mu h_{\lambda\rho}
+ \pd_\nu \pd^\lambda h_{\lambda\mu}
\big)
\vphantom{\frac{h}{2}}
\notag\\
& + \frac{g_{\mu\nu}}{2}
\big(\pd^\lambda \pd^\rho h_{\lambda\rho}
- g^{\lambda\rho} \pd^\sigma \pd_\sigma h_{\rho\lambda}
\big)
\notag\\
& + \frac{h_{\mu\nu}}{2} R
-g_{\mu\nu} h^{\gamma\delta} R_{\gamma\delta}
-\frac{h}{2}G_{\mu\nu}
\notag\\
& + G_{\mu\lambda} {h_\nu}^\lambda
+ G_{\lambda\nu} {h^\lambda}_\mu
+ h^{\gamma\delta}R_{\gamma\nu\delta\mu}
\bigg].\notag
\end{align}
The oscillatory part of this expression is 
\begin{gather}
\widetilde{\left(\frac{\delta \mc{L}_{\rm EH}}{\delta h^{\mu\nu}}\right)}
= \left(\oper{D}^{\rm vac}_{\mu\nu\gamma\delta} + \oper{\mc{G}}_{\mu\nu\gamma\delta}\right) h^{\gamma\delta},
\end{gather}
where the right-hand side is given by \Eqs{eq:dvac} and \eq{eq:mcG}. Hence, one arrives at the results described in \Sec{sec:S2}.



\begin{thebibliography}{10}
\bibitem{foot:mode}
This is similar to electromagnetic waves in plasmas, where transverse modes can
  gradually transform into longitudinal modes when the plasma parameters are
  inhomogeneous \cite[Chap.~13]{book:stix}.
\bibitem{foot:nonloc}
A general theory of dispersive waves in linear media can be found in, for
  example, \Ref{my:nonloc}.
\end{thebibliography}
\end{document}